\documentclass[sigconf]{acmart}
\pdfoutput=1
\usepackage{multirow}
\usepackage{setspace}
\usepackage{array}
\usepackage{subfigure}
\usepackage{tikz}
\usepackage{graphicx}
\usepackage{float}
\usepackage{color}
\usepackage{bbding}
\usepackage[misc]{ifsym}

\AtBeginDocument{%
  \providecommand\BibTeX{{%
    \normalfont B\kern-0.5em{\scshape i\kern-0.25em b}\kern-0.8em\TeX}}}

\setcopyright{acmcopyright}
\copyrightyear{2018}
\acmYear{2018}
\acmDOI{XXXXXXX.XXXXXXX}

\acmConference[MM ’22]{ In Proceedings of the 30th ACM International Conference on
Multimedia }{June 10--14,
  2022}{Lisbon Portugal}
\acmPrice{15.00}
\acmISBN{978-1-4503-XXXX-X/18/06}

\begin{document}

\title{SDRTV-to-HDRTV via Hierarchical Dynamic Context Feature Mapping}


\author{Gang He}
\affiliation{%
  \institution{Xidian University}
  \institution{Kuaishou Technology}
  \streetaddress{}
  \city{}
  \state{}
  \country{}
  \postcode{}
}

\thanks{$^{\textrm{\Letter}}$ Corresponding author.}

\author{Kepeng Xu$^{\textrm{\Letter}}$}
\affiliation{%
  \institution{Xidian University}
  \streetaddress{}
  \city{}
  \country{}}
\email{kepengxu11@gmail.com}

\author{Li Xu}
\affiliation{%
  \institution{Xidian University}
  \city{}
  \country{}
}

\author{Chang Wu}
\affiliation{%
 \institution{Xidian University}
 \streetaddress{}
 \city{}
 \state{}
 \country{}}

\author{Ming Sun}
\affiliation{%
  \institution{Kuaishou Technology}
  \streetaddress{}
  \city{}
  \state{}
  \country{}}

\author{Xing Wen}
\affiliation{%
  \institution{Kuaishou Technology}
  \streetaddress{}
  \city{}
  \state{}
  \country{}
  \postcode{}}
\email{}

\author{Yu-Wing Tai}
\affiliation{%
  \institution{Kuaishou Technology}
  \streetaddress{}
  \city{}
  \country{}}
\email{}

\renewcommand{\shortauthors}{Gang He et al.}



\begin{abstract}
  In this work, we address the task of SDR videos to HDR videos(SDRTV-to-HDRTV conversion). Previous approaches use global feature modulation for SDRTV-to-HDRTV conversion. Feature modulation scales and shifts the features in the original feature space, which has limited mapping capability. In addition, the global image mapping cannot restore detail in HDR frames due to the luminance differences in different regions of SDR frames. To resolve the appeal, we propose a two-stage solution. The first stage is a hierarchical Dynamic Context feature mapping (HDCFM) model. HDCFM learns the SDR frame to HDR frame mapping function via hierarchical feature modulation (HME and HM ) module and a dynamic context \textbf{feature transformation} (DYCT) module. The HME estimates the feature modulation vector, HM is capable of hierarchical feature modulation, consisting of global feature modulation in series with local feature modulation, and is capable of adaptive mapping of local image features. The DYCT module constructs a feature transformation module in conjunction with the context, which is capable of adaptively generating a feature transformation matrix for feature mapping. Compared with simple feature scaling and shifting, the DYCT module can map features into a new feature space and thus has a more excellent feature mapping capability. In the second stage, we introduce a patch discriminator-based context generation model PDCG to obtain subjective quality enhancement of over-exposed regions. PDCG can solve the problem that the model is challenging to train due to the proportion of overexposed regions of the image. The proposed method can achieve state-of-the-art objective and subjective quality results. Specifically, HDCFM achieves a PSNR gain of 0.81 dB at about 100K parameters. The number of parameters is 1/14th of the previous state-of-the-art methods. The test code will be released soon.

\end{abstract}

%
%





\begin{teaserfigure}
  \centering
  \includegraphics[width=0.8\textwidth]{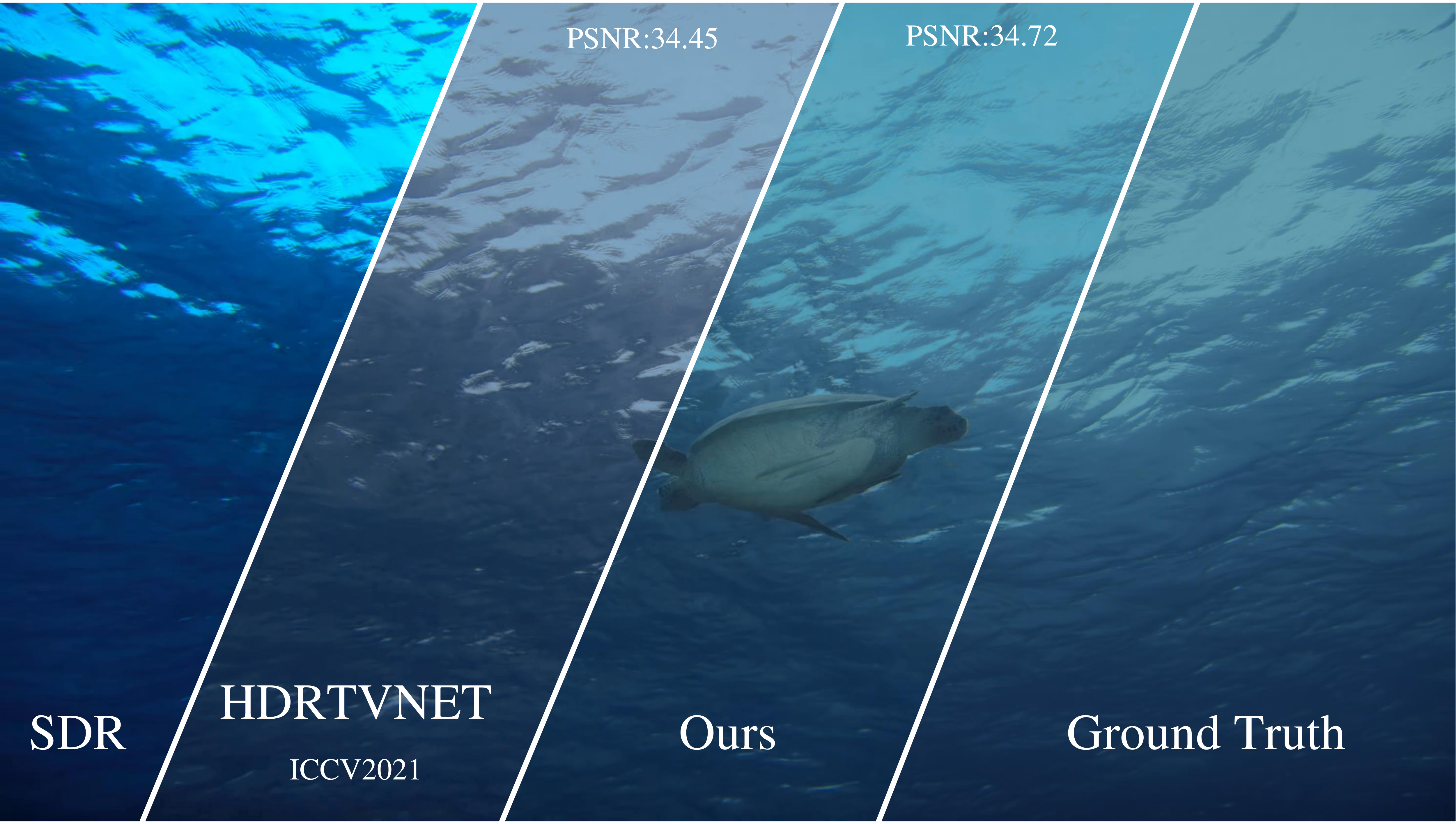}
  \caption{We propose a new deep learning system capable of reconstructing High Dynamic Range (HDR) video from a Standard Dynamic Range (SDR) video. We introduce hierarchical global and local feature modulation, which allows different processing of different local image. And we introduce a model of local feature transformation that can model stronger feature mapping. As can be seen from the figure, the proposed method is able to generate HDR frames that are closer to the ground truth. All images have not been additionally processed to preserve all detail of the HDR frames, an HDR display is required to fully display the visual quality of HDR frames, and playback on an SDR display will be dark.}  
  \label{Fig1}
\end{teaserfigure}

\maketitle

\begin{figure}[h]
  \centering
  \setlength{\abovecaptionskip}{0.cm}
  \vspace*{-12pt}
  \includegraphics[width=0.49\textwidth]{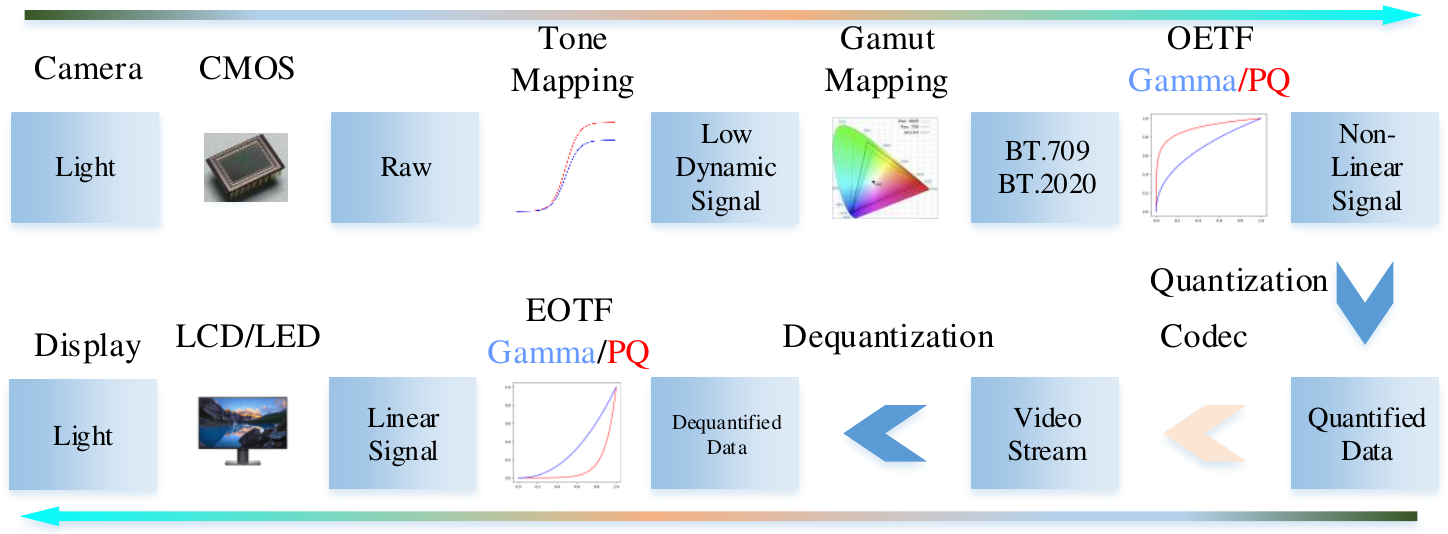}
  \caption{\textcolor{blue}{SDR}/\textcolor{red}{HDR} video processing framework from the capture side to the playback side. SDR and HDR use different settings in Tone mapping, Gamut Mapping, OETF and EOTF stages.}

  \label{SDR2HDR}
  \setlength{\belowcaptionskip}{-0.cm}
  \vspace*{-17pt}
\end{figure}

\section{Introduction}

High Dynamic Range (HDR) video allows for a more realistic display and reproduction of the natural world. HDR video has a higher bit depth, wider color range, and higher brightness per pixel.
Although HDR display device technology is now widely available and HDR video has many advantages, most video sources are still stored in Standard Dynamic Range (SDR) format.
Therefore, converting many existing SDR videos to HDR videos can dramatically improve the user experience.

We present the process of SDR/HDR video content acquisition and playback in Fig.\ref{SDR2HDR}.
From the moment light enters the camera to the playback of the video image using the monitor, it goes through the following stages:
1) convert the light signal into a digital signal through CMOS;
2) reduce the high dynamic digital signal to a low dynamic signal by Tone Mapping\cite{mantiuk2008display}; 
3) transfer the image color to the target color gamut by Gamut mapping;
4) convert linear signal to nonlinear signal by OETF \cite{PQ,HLG} ;
5) quantize digital signal and arithmetic coding\cite{rissanen1979arithmetic};
6) encode and decode by codec\cite{8456249,H265};
7) convert the decoded nonlinear signal to a linear digital signal by EOTF;
8) convert the linear signal to an optical signal for playback.
The main difference between SDR and HDR is using different EOTF(Electro-Optical Transfer Function) and OETF(Optical-Electro Transfer Function).

To distinguish SDR video to HDR video from the SDR image to HDR image task, we follow the definition of method \cite{HDRTVNET} and define SDR video to HDR video as SDRTV-to-HDRTV conversion.
LDR-to-HDR (LDR image to HDR image) refers to the conversion of SDR image to HDR image. HDR images can play on display devices by tone mapping.
The previous approach \cite{SRITM,JSIGAN} united super-resolution with SDRTV-to-HDRTV conversion, and tried to build a pipeline from low-resolution SDR video to high-resolution HDR video.
HDRTVNET\cite{HDRTVNET} proposed a multi-stage scheme to complete SDRTV-to-HDRTV conversion by global tone mapping, local image enhancement, and image generation.

In the SDRTV-to-HDRTV conversion, the most crucial issue is to map the SDR features to the HDR feature domain, which is called feature mapping in this paper.
The second issue is generating information on over-exposure areas that do not exist in SDR.
An SDR video to HDR video conversion pipeline is constructed to address these two issues.
The pipeline is divided into two parts,  
Hierarchical Dynamic Context Feature Mapping model (HDCFM) and a Patch discriminator-based Dynamic Context Generation network(PDCG). 
The first part obtains HDR frames with superior objective quality by feature mapping, and the second part accomplishes over-exposure area image enhancement.

Specifically, HDCFM contains the Hierarchical feature Modulation vector Estimation (HME) module, Hierarchical Modulation (HM) module, and Dynamic Context feature Transformation (DYCT) module.
For HME, we construct a hierarchical modulation vector estimation module that captures the global and local image prior to estimating the global and local feature modulation vectors.
For HM, the global and local modulation vectors estimated by HME are used to modulate the input features.
Such feature modulation enables adaptive mapping of local images in different regions of different frames.
For DYCT, we propose the joint context local feature transformation module to extract image context information and accomplish local feature transformation by dynamic convolution. HDCFM can complete feature mapping and obtain HDR frames based on the above structure.
The proposed HDCFM has two advantages over the previous methods. 
The first is that the proposed HM and HME can perform spatially adaptive mapping using image local information. 
In addition, the proposed DYCT module models a more robust dynamic feature transformation: the ability to transform features directly to a new feature space instead of the previous simple feature scaling and shifting. 
This dramatically enhances the mapping performance of the model.
Thus, a more complex mapping process can be modeled to map SDR frames to HDR frames better.
For PDCG, a Patch GAN with an over-exposure mask is proposed, which can generate over-exposure region image information to obtain the higher subjective quality of HDR frames.
The proposed HDCFM with a smaller number of parameters can obtain the best conversion performance.
In order to compare with previous methods, we selected five evaluation metrics PSNR, SSIM, SR-SIM\cite{SRITM}, $\Delta E_{ITP}$ and HDR-VDP3\cite{mantiuk2011hdr} to evaluate the proposed method.

In summary, our contributions include the following main points.

\begin{itemize}
  \vspace{-5pt}
  \item We propose a hierarchical feature modulation module that can perform spatially adaptive feature modulation on image features; local feature modulation can improve the quality of HDR reconstructed frames.

  \item We propose a dynamic feature transformation method that can further improve the feature mapping capability of the model to obtain higher quality HDR converted frames.

  \item We analyzed the problem of over-exposure in the SDRTV-to-HDRTV conversion. Propose a Patch discriminator-based over-exposure image generation model that can obtain a higher subjective quality HDR frame.

  \item With about 100K parameters, the proposed method can obtain state-of-the-art results compared to previous methods. 
  \vspace{-20pt}
\end{itemize}

\section{Related work}

Converting previous SDR videos to HDR videos is a valuable task. More and more researchers are focusing on this topic.
There are several main methods for SDR to HDR conversion as follows.
1) Multi-exposure SDR images to single-frame HDR images.
2) single-frame SDR image to single-frame HDR image.
3) SDR video to HDR video.
Our goal is to convert SDR video that already existed to HDR video.

\textbf{LDR-to-HDR}. The traditional method estimates the light source density, based on which the dynamic range is further expanded \cite{1111,2222,3333,4444}. 
Researchers have proposed a method based on deep convolutional neural network \cite{liu2020single} to convert LDR images to HDR images directly.
HDRCNN\cite{HDRCNN,liu2020singleimage,3392403} propose method that can recover the over-exposure area of the image.
\cite{drhdri,drhdri,5555,yan2020deep,niu2021hdr} proposed methods can predict multi-exposure LDR image pairs by a single LDR image, then synthesize HDR images based on the predicted multi-exposure image pairs.

\textbf{SDRTV-to-HDRTV conversion}. The SDRTV-to-HDRTV conversion approach has only emerged in the last two years.
\cite{SRITM} proposes a GAN-based architecture that jointly achieves super-resolution and SDTV to HDRTV.
\cite{JSIGAN} proposes a hierarchical GAN architecture to accomplish super-resolution and SDRTV  to HDRTV.
\cite{HDRTVNET} proposed a method using global feature modulation, local enhancement, and over-exposure compensation, which achieved the best performance.

\textbf{Dynamic Convolution}
The vanilla convolutional layer learns the parameters of the convolutional kernel through big data during training.
The parameters of the convolution kernel are fixed during the inference phase and do not change for different inputs; such a convolution kernel is also called a static convolution kernel.
Dynamic convolution, meaning that the convolution kernel parameters are dynamically updated during the inference phase, allows the model to extract more complex features and build complex pattern recognition methods.
Current researchers focus on how to construct dynamic convolution kernels \cite{debrabandere2016dynamic,zhou2021decoupled}.

\textbf{Context Convolution}
In convolutional neural networks, context extraction aims to extract the correlation between the current location features and the global features, thus improving the modeling capability.
To extract non-local information from images \cite{wang2018nonlocal} proposes a non-local generic module to complement the long-range dependencies to capture the global context information.
\cite{gcnet} finds that the global relevance information obtained from different locations during non-local modeling is almost the same.
Therefore, a generic feature aggregation module is proposed to extract global feature information directly instead of global relevance for each feature element.
Such a modeling approach dramatically reduces the computational cost and improves the feature extraction performance.

\textbf{Patch GAN}
The Generative Adversarial Networks (GAN) model \cite{goodfellow2014generative,metz2017unrolled,stylegan,tolosana2020deepfakes,wang2020cnn,gui2021review} has been widely used in the field of image generation in recent years, thanks to its unique architecture design.
Patch GAN\cite{patchgan} improves the clarity of the generated images, enabling higher resolution images.
In this paper, the Patch discriminator can solve the problem of low model performance caused by the low proportion of over-exposure region.

\begin{figure}[t]
    \centering
    \vspace*{-5pt}
    \includegraphics[width=0.49\textwidth]{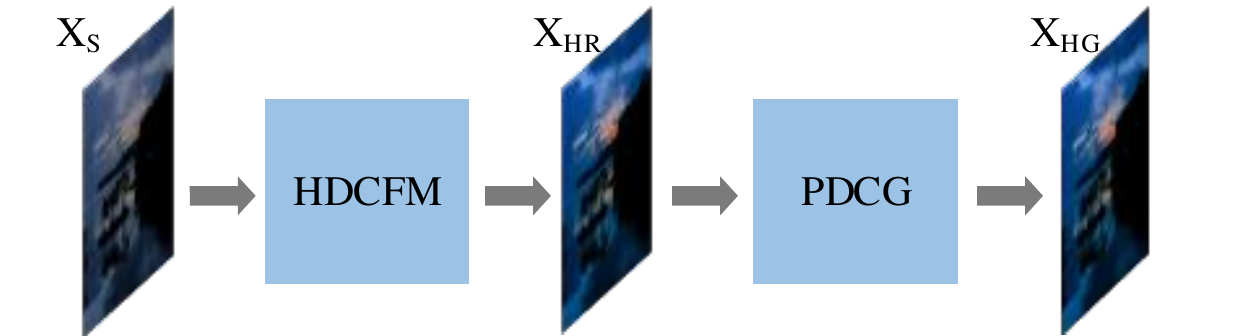}
    \caption{The framework of the proposed method. 
    Firstly, a Hierarchical Dynamic Context Feature Mapping model(HDCFM) Convert Standard Dynamic Range(SDR) frame $X_S$ to High Dynamic Range(HDR) frame $X_{HR}$.
    HDCFM consists of a Hierarchical Modulation Estimation module(HME), a Hierarchical feature Modulation module(HM) and a Dynamic Context feature Transformation module(DYCT). 
    Secondly, we propose a Patch discriminator-based Dynamic Context Generation model(PDCG). 
    PDCG enhances the over-exposed region in the HDR frame $X_{HR}$ output from HDCFM, resulting in a subjectively higher quality HDR frame $X_{HG}$.
    }
    
    \label{Framework}
    \vspace*{-10pt}
\end{figure}

\begin{figure*}[h]
  \centering
  \includegraphics[width=0.98\textwidth]{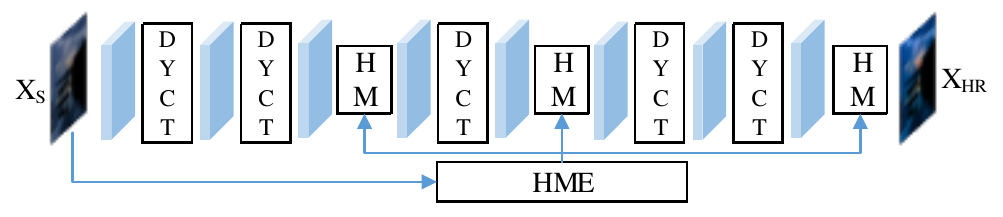}
  \caption{The architecture of Hierarchical Dynamic Context Feature Mapping (HDCFM). In HDCFM, firstly, the SDR frame $X_S$ is input to the HME module and the hierarchical feature modulation vector is obtained. Next, $X_S$ is passed through the DYCT and HM modules to obtain the HDR reconstruction $X_{HR}$.}
  \label{HDCFM}

\end{figure*}
\section{Methodology}

\subsection{Framework}

SDR frames to HDR frames can be modeled as a feature mapping and feature complement process.
For this purpose, we propose a two-stage solution. The first stage converts SDR frames $X_S$ to HDR frames $X_{HR}$ by a Hierarchical Dynamic Context Feature Mapping model HDCFM. The second stage uses a Patch discriminator-based Dynamic Context Generation model PDCG to complete the over-exposure enhancement and obtain HDR frames $X_{HG}$ with higher subjective quality.
The framework of the whole scheme is shown in Fig.\ref{Framework} and Formula (\ref{FormulasAll}).
The main challenge of the SDRTV-to-HDRTV conversion is that the data distribution of SDR frames differs significantly from that of HDR frames and that SDR frames store less information(There are overexposure problems).
$M_H$ is the over-exposure area mask calculated similarly as \cite{Marcel:2020:LDRHDR}.
The specific motivation and architecture of each module will be described next.

\begin{equation}
	\begin{aligned}
		X_{HR}&=HDCFM(X_S) \\
		X_{HG}&=PDCG(X_{HR},M_H) 
	\end{aligned}
\label{FormulasAll}
\end{equation}

\begin{figure}[h]
  \vspace*{-25pt}
  \centering
  \includegraphics[width=0.45\textwidth]{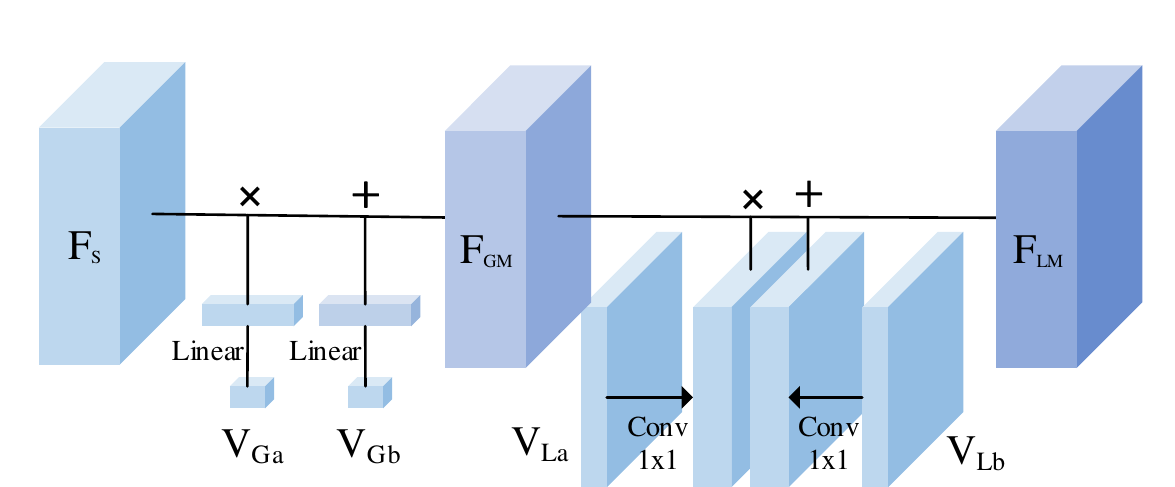}
  \caption{The architecture of Hierarchical Modulation (HM). The global and local feature modulation tandem constitutes the HM.}
  \label{HM}
  \vspace*{-25pt}
\end{figure}
\subsection{HDCFM}

In this part, to transform SDR frame $X_S$ to HDR frame $X_{HR}$, we construct HDCFM, which can achieve the instance-space adaptive feature mapping method in the image feature space. Specifically, HDCFM can perform different feature mapping for different input and pixels at different spatial locations in the input image. A locally adaptive image feature mapping model is constructed, which can obtain high quality HDR frames.

The critical point 1 of HDCFM is the hierarchical feature modulation vector estimation module HME and the hierarchical feature modulation module HM, which can accomplish hierarchical global and local feature modulation.
Essentially, the SDR image features are scaled and shifted in the feature space to obtain features closer to HDR images.
The critical point 2 of HDCFM lies in the dynamic feature transformation module DYCT of context features, which can accomplish dynamic feature transformation (matrix transformation). 
The structure of HDCFM is shown in Fig.\ref{HDCFM}.
The motivation and methodology of these two points will introduce next.

\subsubsection{Motivation of Feature Mapping}

The feature extracted from SDR video frame is in SDR feature space, and the feature extracted from HDR video frame is in HDR feature space, so SDRTV- to-HDRTV conversion can be modelled as a feature mapping. For the input SDR frame $X_S$, the low dynamic feature $F_S$ is first obtained by convolution, and then $F_S$ needs to be mapped to the high dynamic feature $F_H$, and finally $F_H$ is recovered to the image space.
The feature mapping proposed in this paper consists of two parts, which are hierarchical feature modulation and local feature transformation.

\subsubsection{Motivation of HM}

During SDRTV-to-HDRTV conversion, pixels at different spatial locations need to be processed differently. For example, in one frame, there are both over-exposed and under-exposed areas, then different processing should be performed on the both under-exposed and over-exposed image area. To address this problem, we design HM composed of global feature modulation and local feature modulation. The global feature modulation can make macro adjustments to the image, and the local feature modulation can further complete the local fine-tuning.

\subsubsection{Architecture of HM}
\begin{figure}[h]
  \vspace*{-10pt}
  \centering
  \includegraphics[width=0.45\textwidth]{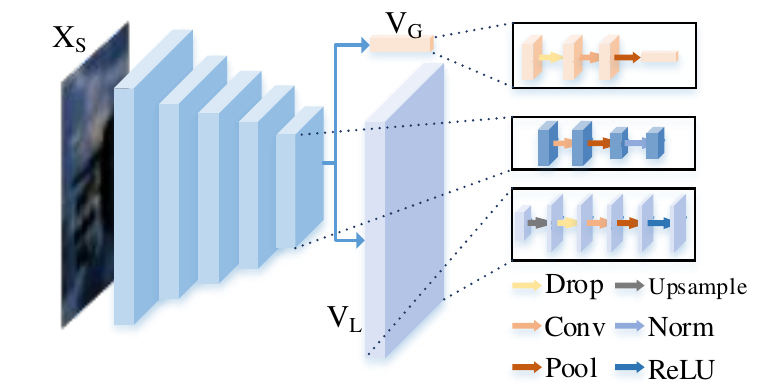}
  \caption{The architecture of Hierarchical Modulation Estimation(HME) module.}
  \label{HME}
  \vspace*{-10pt}
\end{figure}

To obtain spatially adaptive feature modulation vectors, our HDCFM constructs a joint global and local hierarchical modulation vector estimation module HME. HME can predict not only the global feature modulation vectors $V_{Ga}$, $V_{Gb}$, but also the local feature modulation vectors $V_{La}$, $V_{Lb}$; this gives the ability to perform different feature modulations on image features at different spatial locations.
The structure of HM is shown in Fig.\ref{HM}. 
Then comes calculating the feature modulation vector by the HME module.
Specifically, for the input $X_S$, $F_{D5}$ is obtained after five downsamples.
$F_{D5}$ goes through the global downsample to obtain $V_{Ga}$ and $V_{Gb}$.
$F_{D5}$ passes through the upsample to obtain $V_{La}$ and $V_{Lb}$.
Such a calculation process is shown in Fig.\ref{HME}.
Next, the feature modulation of $F_S$ is performed using HM.
The HM is divided into two steps.
The first step uses the global modulation parameter $V_{Ga}$ to dot-multiply $F_S$, followed by adding the features after point multiplication using $V_{Gb}$. The second step uses the local modulation parameter $V_{La}$ to dot-multiply $F_S$, followed by $V_{Lb}$ to sum the dot-multiplied features.

\subsubsection{Motivation of the DYCT Module}.
The core of DYCT (Dynamic Context Feature Transformation module) is feature transformation.
Firstly, we introduce the difference between feature modulation and feature transformation.
\textbf{Feature modulation}: The modulated feature layer is obtained by dot multiplying the feature layer $F_S$ by the modulation vector, a simple feature mapping that only scales and shifts the features in the original feature space. So it can model a relatively single mapping process. This modeling process is shown in Fig.\ref{HM}.
\textbf{Feature transformation}: feature transformation (matrix multiplication) of the input features $F_S$ can transform $F_S$ from the original feature space to the new feature space. This modeling approach has a much more robust feature mapping capability.
Therefore, we propose to build a feature mapping module based on feature transformation. It can model a more complex feature mapping process and thus able to obtain better HDR transformed frames.

\subsubsection{Modeling Process of the DYCT Module}
\begin{figure}[h]
  \centering
  \includegraphics[width=0.48\textwidth]{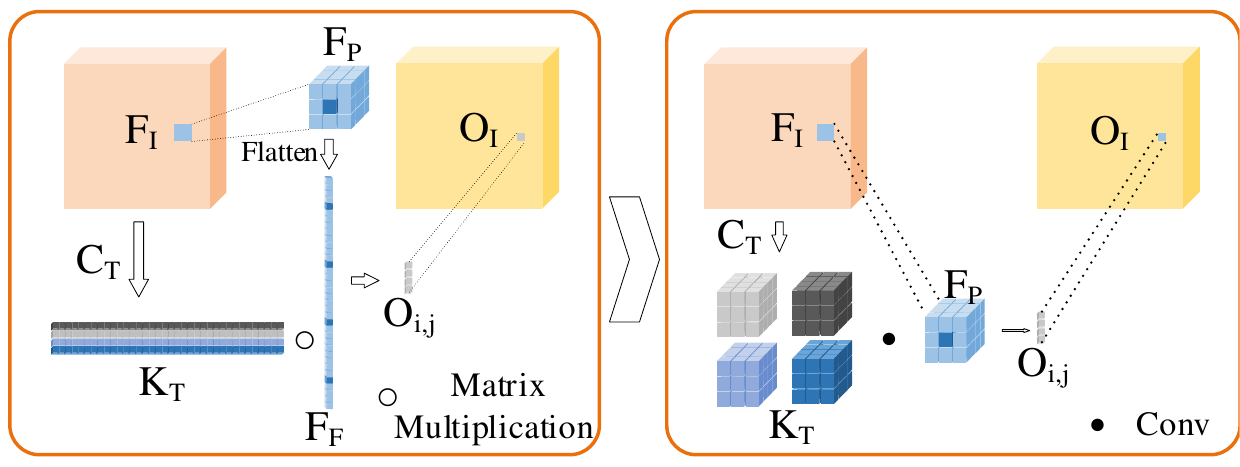}
  \caption{The conceptual architecture of the local feature transform model, on the left is the matrix multiplication implementation of the local feature transform. For the input feature map, when processing to the local feature $F_P$. We use the feature transformation to obtain $O_{i,j}$.  Here we need a feature transformation matrix $KS$, so we predict this feature transformation matrix $KS$ and use $KS$ to matrix multiply with $F_P$ to get $O_{i,j}$.
  On the right is a dynamic convolution implementation of this operation. These two implementations are exactly equivalent in a mathematical sense. Modern deep learning frameworks are more optimized for convolution, so we use dynamic convolution to accomplish the local feature transform.}
  
  \label{LLT}
\end{figure}

.
We begin with an introduction to the local feature transformation module, as shown in the left part of Fig.\ref{LLT}, for the local feature $F_P$ (with shape $(K,K,C_I)$) of input $F_S$, is flattened to a vector $F_F$ of ($K\bullet K\bullet C_I,1$).
At the same time, we need a conditional generation module $C_T$ to predict the parameters $K_T$ of the feature transformation (in shape $(C_O, K\bullet K\bullet C_I)$).
Next, $K_T$ is subjected to a matrix multiplication operation with $F_F$, which in algebra is called the linear feature transformation, and finally, the transformed feature $O_{i,j}$ (with shape $(C_O, 1)$) is obtained.

To further analyze the process of feature transformation, we can use the local feature transformation in convolution
specific implementation form.
As shown in the right Fig.\ref{LLT}, for the input local feature $F_P$, $C_O$ convolution kernels $K_T$ of shape $(K,K,C_I)$ are needed to convolve with $F_P$. The output is $O_{i,j}$, where $K_T$ is generated online by the $C_T$ module when inference is needed. $C_I$ is the number of channels in the input feature layer.
We continue to analyze the generation of $K_T$. In a practical application, the resolution of the input image is ($H\bullet W$), and the size of the input feature layer of the local feature transform is ($H,W,C_I$).
For each pixel, $C_O$ convolution kernels are predicted, and the shape of each convolution kernel is $(K,K , C_I)$. The number of parameters for all convolution kernels is ($H \bullet W \bullet C_O \bullet K \bullet K \bullet C_I$ ). The total number of parameters in the 4K image processing task is $3.057 \times 10^{11}$, directly leading to memory out.

\begin{table*}
  \begin{spacing}{1.05}
  \vspace*{-10pt}
  \centering
  \caption{Quantitative comparisons. {\color[HTML]{FE0000} \bf{Red}} text indicates the best and {\color[HTML]{3166FF} blue} text indicates the second. The result of the previous methods in the table are consistent with \cite{HDRTVNET}.}
  \begin{tabular}{m{4cm}<{\centering}m{1.7cm}<{\centering}m{1.7cm}<{\centering}m{1.7cm}<{\centering}m{1.7cm}<{\centering}m{1.7cm}<{\centering}m{1.7cm}<{\centering}}
  \toprule
  Methods & Params$ \downarrow$ & PSNR$ \uparrow$  & SSIM$ \uparrow$  & SR-SIM$ \uparrow$  & $\Delta E_{ITP} \downarrow$  & HDR-VDP3$ \uparrow$  \\ 
  \midrule
  HuoPhyEO\cite{huo2014physiological}TVC & - & 25.90 & 0.9296 & 0.9981 & 38.06  &  7.893 \\ 
  Kovaleski\cite{kovaleski2014high}\scriptsize SIBGRAPI & - & 27.89 & 0.9273 & 0.9809 & 28.00 & 7.431  \\
  ResNet\cite{he2016identity}ECCV16 & 1.37M & 37.32 &0.9720& 0.9950 & 9.02 & 8.391  \\
  Pixel2Pixel\cite{isola2017image}CVPR17 & 11.38M & 25.80 & 0.8777 & 0.9871 & 44.25 & 7.136  \\
  CycleGAN\cite{zhu2017unpaired}ICCV17 & 11.38M & 21.33 & 0.8496 & 0.9595 & 77.74 & 6.941  \\
  HDRNET\cite{gharbi2017deep}TOG & 482K & 35.73 & 0.9664 &0.9957 & 11.52 & 8.462 \\
  CSRNET\cite{he2020conditional}ECCV20 & \textcolor{red}{\bf{36K}} & 35.04 & 0.9625 & 0.9955 & 14.28 & 8.400  \\
  Ada-3DLUT\cite{zeng2020learning}TPAMI & 594K & 36.22 & 0.9658 & \textcolor{blue}{0.9967} & 10.89 & 8.423 \\
  Deep SR-ITM\cite{kim2019deep}ICCV19 & 2.87M & 37.10 & 0.9686 & 0.9950 & 9.24 & 8.233  \\
  JSI-GAN\cite{kim2020jsi}AAAI20 & 1.06M & 37.01 & 0.9694 & 0.9928 & 9.36 & 8.169  \\
  HDRTVNET\cite{HDRTVNET}ICCV21 & 1.41M & \textcolor{blue}{37.61} & \textcolor{blue}{0.9726}  & \textcolor{blue}{0.9967} & \textcolor{blue}{8.89} & \textcolor{red}{\bf{8.613}}  \\
  HDCFM(Proposed) & \textcolor{blue}{100.63K} & \textcolor{red}{\bf{38.42}} & \textcolor{red}{\bf{0.9732}}  & \textcolor{red}{\bf{0.9974}} & \textcolor{red}{\bf{7.83}}  &  \textcolor{blue}{8.5716}  \\ 
  \bottomrule
  \label{Quantitativetable}
  \end{tabular}
  \end{spacing}
  
  \vspace*{-15pt}
\end{table*}

\begin{equation}
    \begin{aligned}
        K_S&=SKP(F_{S}) \\
        K_C&=CKP(F_{S}) \\
        F_{mid}&=DDF(F_S,K_S,K_C) \\
        F_O&=CB(F_{mid}) \\
        CKP&=GAP \bullet Conv \\
        SKP&=Conv \bullet Conv \\
        CB&=SpatialConv \bullet  ChannelConv 
        \label{DCTFORMULA}
    \end{aligned}
\end{equation}

\subsubsection{Architecture of the DYCT Module}
\begin{figure}[h]
    \centering
    \includegraphics[width=0.5\textwidth]{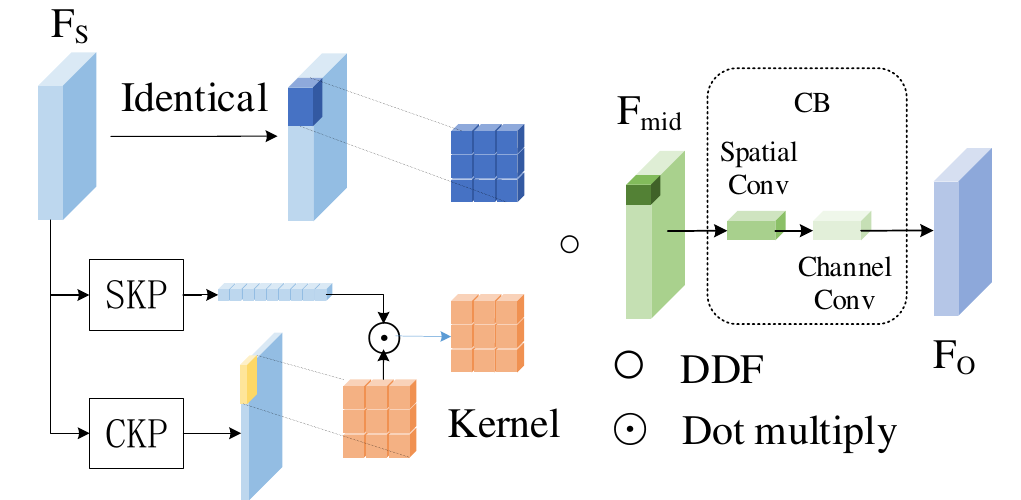}
    \caption{The architecture of Dynamic Context feature Transformation(DYCT). The input features $F_S$ are input to the SKP and CKP modules to predict the convolution kernel parameters (feature transformation matrix) and further perform feature transformation on $F_S$ to obtain the transformed features. This is able to model a more complex feature mapping process than feature modulation.}
    \label{DCTmodule}
\end{figure}

.
This paragraph will introduce the specific architecture of the DYCT. To solve memory out, we borrow the idea of decoupled dynamic convolution kernels.
The architecture of the whole DYCT module is shown in Fig.\ref{DCTmodule}.
The specific process is as follows.
Decompose the original $C_O$ convolution kernel $K_T$ into a combination of spatial convolution kernel $K_S$(K,K,H,W) and channel convolution kernel $K_C$(C,K,K), $K_S$ and $K_C$ generated by $SKP$ and $CKP$ respectively, the computation process of $SKP$ and $CKP$ is define in Formula(\ref{DCTFORMULA}).
After obtaining $K_S$ and $K_C$, the output feature layer $F_{mid} $ is obtained by convolving through the decoupling convolution method $DDF$ proposed by \cite{zhou2021decoupled}.
During the convolution of dynamic filters, the convolution kernel weights are obtained in real-time by sample inference. This can enhance the feature mapping capability of the model.
This is a convolutional implementation of the feature transformation.

It is worth noting that only local feature transformations may lead to large differences in the transformed results for image contents that have the same color in different regions.
Therefore, this paper uses a context module to count the overall feature context information and fine-tune the features.
In this part, we choose the context information module $CB$ proposed by \cite{gcnet}, input $F_{mid} $ to $CB$, and the output result is the final output of the DYCT module.
The structure of $SpatialConv$ and $ChannelConv$ is similar to \cite{gcnet}, using DDF convolution instead of vanilla convolution. 
This module has the advantage of being computationally small, and the computation procedure is given in Formula(\ref{DCTFORMULA}).

\subsection{PDCG}

\begin{figure}[h]
  \setlength{\abovecaptionskip}{0.cm}
    \centering
    \includegraphics[width=0.5\textwidth]{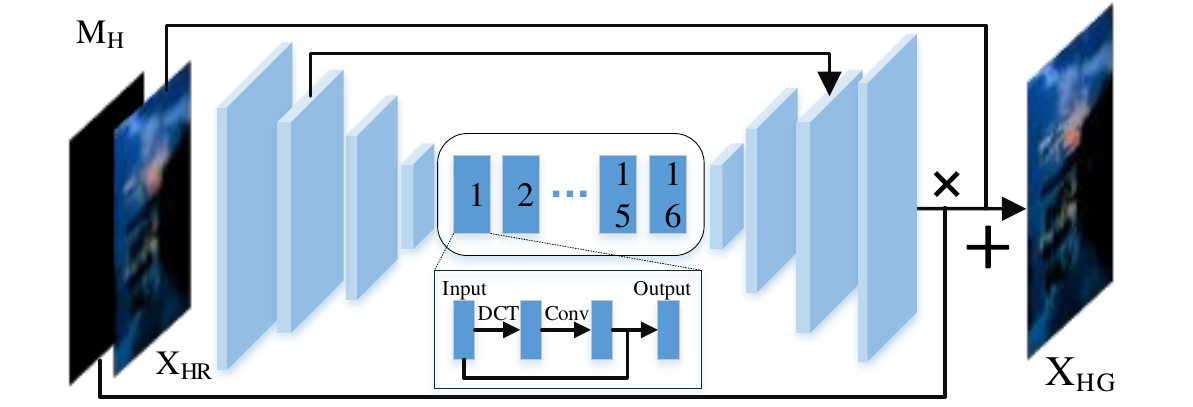}
    \caption{The architecture of Patch discriminator Dynamic Context Generation model (PDCG).}
    \label{pdcg}
    \setlength{\belowcaptionskip}{-0.cm}
\end{figure}

The objective quality of $X_{HR}$ generated by HDCFM is very high, but $X_{HR}$ still suffers from the problem of missing over-exposure information. 
The problem of loss of over-exposure information can cause large bright spots or even wrong colors in HDR reconstructed frames.
To be able to generate over-exposure region image information, we propose the PDCG model, the structure of PDCG is shown in Fig. \ref{pdcg}.
The generation method of the over-exposure section is formally defined in Formula(\ref{hg}), and the mask of the over-exposure section is defined as $M_H$.
We input the preliminary results $X_{HR}$ generated by HDCFM into the PDCG model.
After three convolutions with a stride of 2, the resolution of the feature layer is reduced, and $F_{d1}, F_{d2}, F_{d3}$ are obtained.
$F_{d3}$ is input into 16 blocks in series.
Each block contains a DYCT module, a vanilla convolution, and a skip connection.
Then perform an upsampling to obtain $F_{u1}$, add $F_{u1}$ and $F_{d2}$, and continue to upsample twice to obtain the final $X_{HG}$. The entire architecture of PDCG is shown in Fig.\ref{pdcg}.
We use a loss function $L_{HG}$ with $L1$ loss, perceptual loss $L_P $, and adversarial loss $L_{GAN}$ combined. The definition of joint loss $L_{HG}$ is shown in Formula(\ref{FLHG}), and $\alpha$,$\beta$ and $\gamma$ are taken as 1.0, 0.5, 0.005 respectively.
We use the pre-trained VGG19\cite{vgg19} on ImageNet1000\cite{russakovsky2015imagenet} to compute the perceptual loss, which improves the subjective quality of the reconstructed frames.
Since the perceptual loss is more dependent on the model structure \cite{liu2021generic}, the model trained on Imagnet to compute the perceptual loss of HDR video frames is still valid.
Since the percentage of highlight regions is deficient, we use Patch-based adversarial loss to generate realistic over-exposure image.

\begin{equation}
    X_{HG}=PDCG(X_{HR},M_H) \bullet M_H + X_{HR} \bullet (1-M_H)
    \label{hg}
\end{equation}

\begin{equation}
    L_{HG} = \alpha  L_1 + \beta  L_P + \gamma  L_{GAN}
    \label{FLHG}
\end{equation}

\section{Experiment}

\begin{figure*}[t]
    \centering
    \setlength{\abovecaptionskip}{0.cm}
    \includegraphics[width=0.95\textwidth]{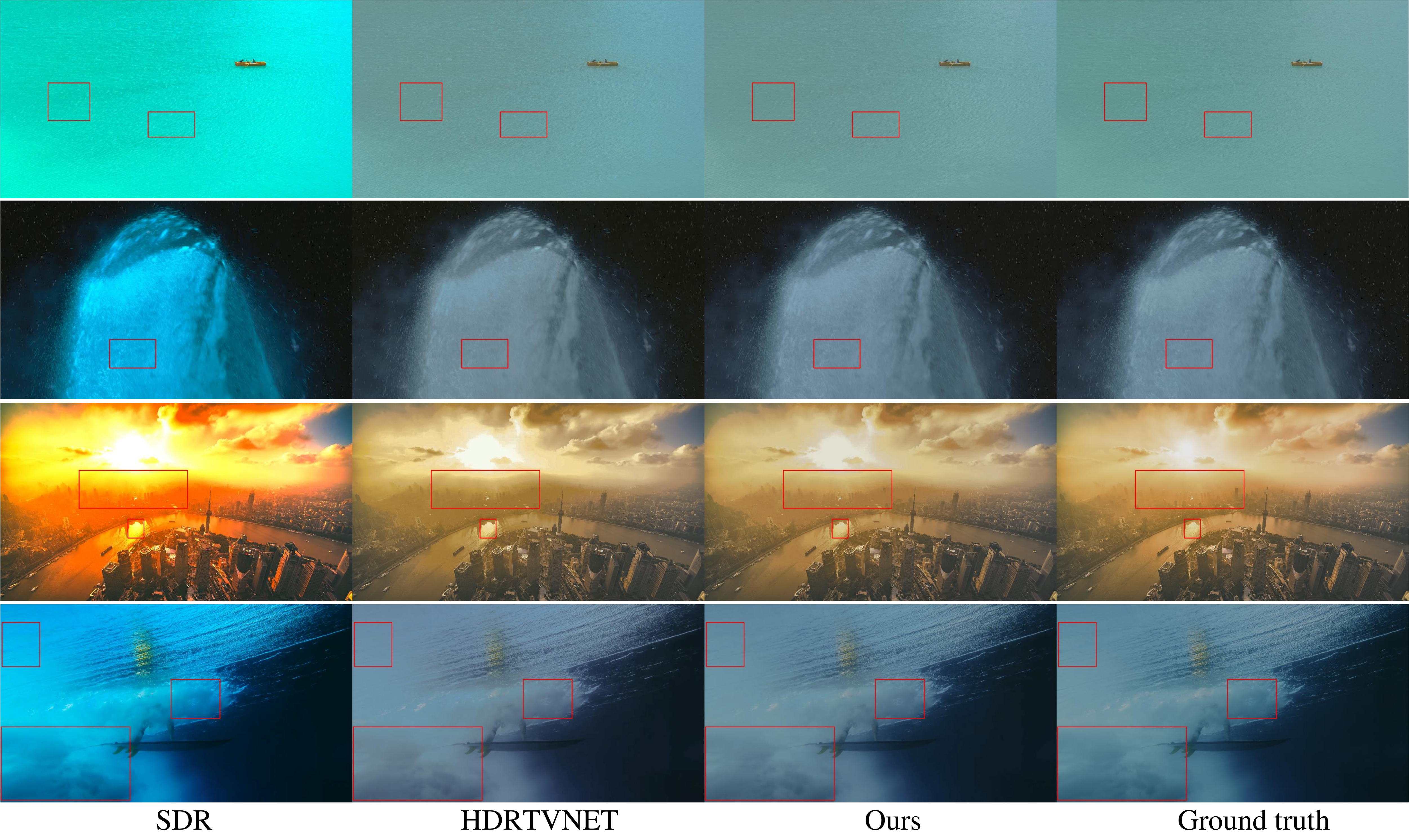}
    \caption{HDR reconstruction of frame results. The proposed method is able to reconstruct local color information in higher quality compared to previous state-of-the-art methods. The proposed method in the red box clearly achieves higher quality color conversion results.}
    \setlength{\belowcaptionskip}{-0.cm}
    \label{QualitativeImage1}
\end{figure*}

\begin{figure}[t]
  \setlength{\abovecaptionskip}{0.cm}
    \centering
    \includegraphics[width=0.47\textwidth]{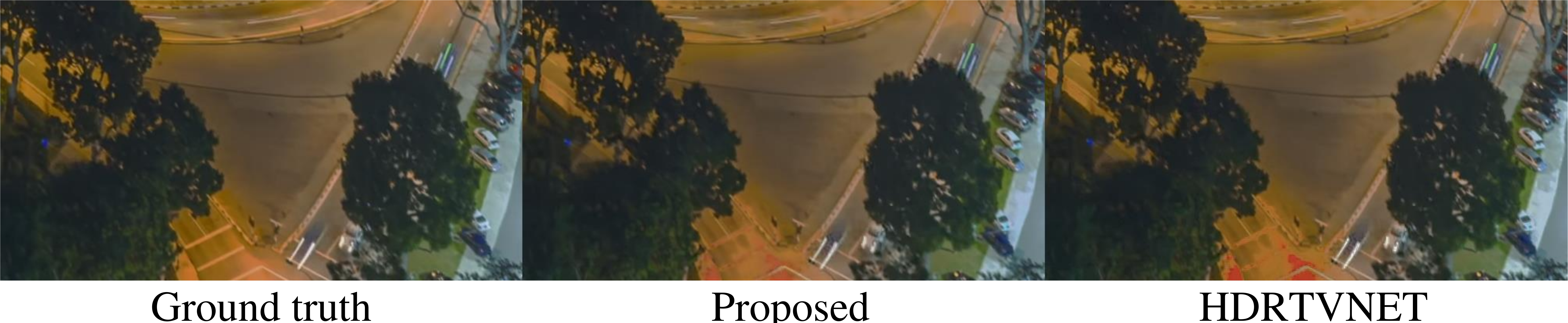}
    \caption{Local results. The proposed HDCFM can perform adaptive feature mapping based on local information, thus effectively improving the quality of HDR reconstructed frames in overexposed regions.}
    \label{QualitativeImage2}
\end{figure}

\subsection{Experiment Settings}

\textbf{Dataset}. For a fair comparison with previous methods, we use the dataset used by \cite{HDRTVNET} captured HDR and SDR versions of each video; each HDR video was HDR10 with BT.2020 color gamut. Frames from the videos were extracted using FFMPEG, cropping the images to 480x480 size.
117 pairs of unduplicated images were included in the test set, each at 4K in size.

\textbf{Training Setup}. In the model's training process, we use L1 as the loss function to optimize the HDCFM model.
The Adam optimizer is used, the initial learning rate is set to 0.0005, and the learning rate is set to 1/2 of the initial rate every 200000 iterations; the total number of iterations is set to 1000000.

\textbf{Evaluation Setup}. To verify the effectiveness of the proposed method, we evaluate the effectiveness of the proposed method on the evaluation index of PSNR, SSIM, SR-SIM , $\Delta E_{ITP}$  and HDR-VDP3.


\begin{figure}[h]
  \setlength{\abovecaptionskip}{0.cm}
  \centering
  \includegraphics[width=0.45\textwidth]{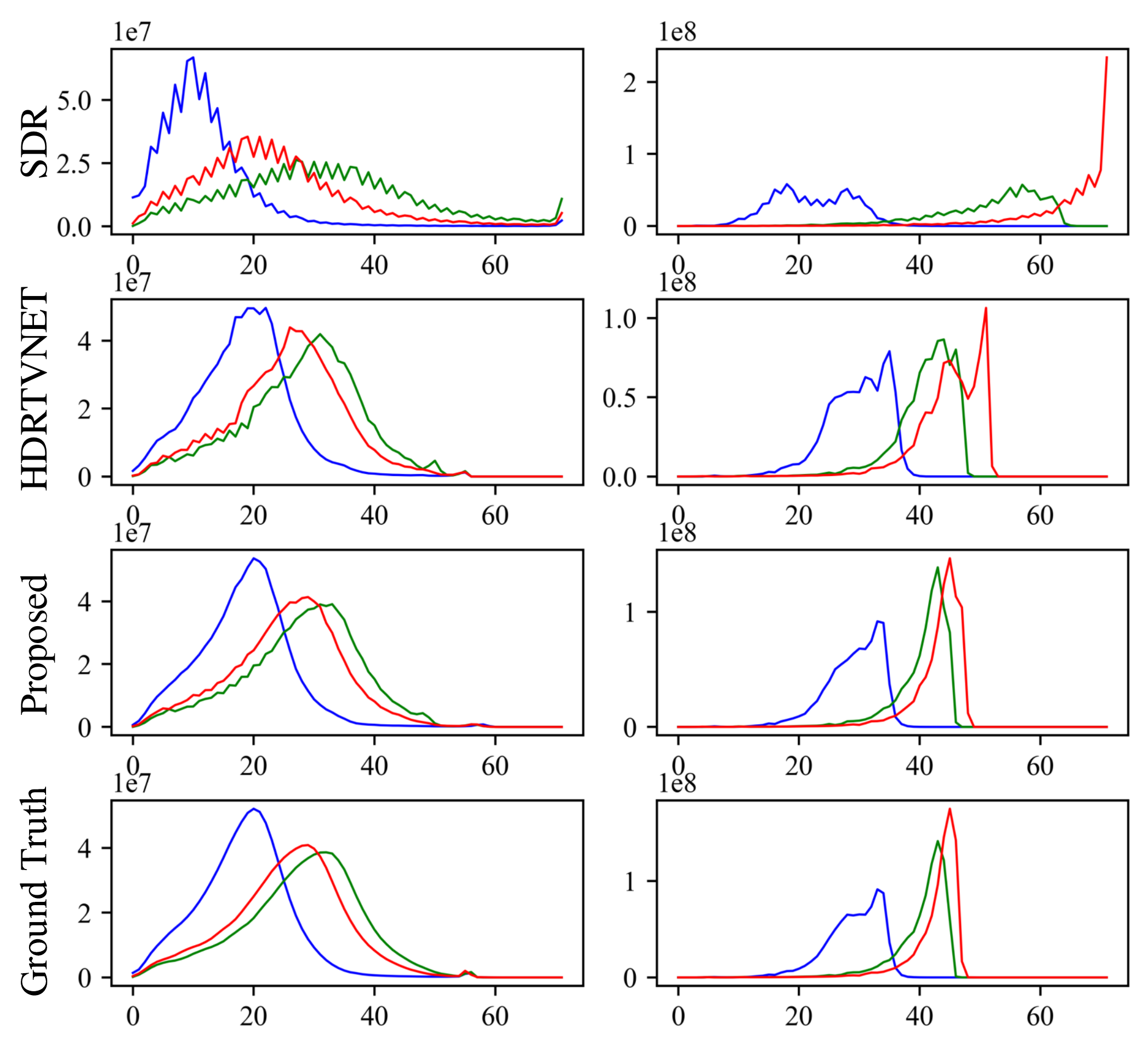}
  \caption{We analyze the histograms of HDR video frames generated by different methods. 
  The left and right are the histograms of two different frames obtained by different methods. 
  The histogram density is set to 72, and the proposed method can obtain smoother and more accurate histogram results compared to the previous methods.}
  \label{hist}
  \setlength{\belowcaptionskip}{-0.cm}
\end{figure}

\begin{figure}[h]
  \setlength{\abovecaptionskip}{0.cm}
  \centering
  \includegraphics[width=0.48\textwidth]{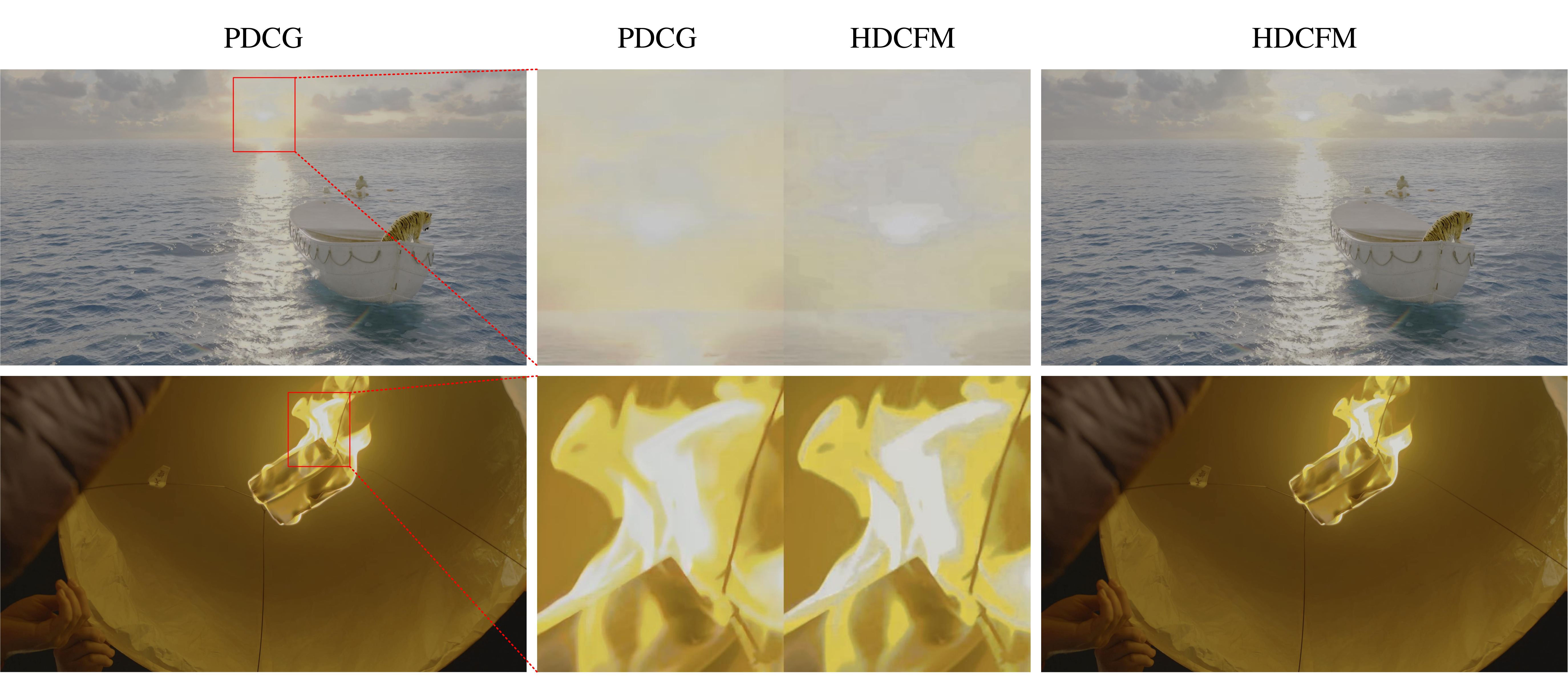}
  \caption{Qualitative results of Patch discriminator-based Dynamic Context Generation model(PDCG). PDCG can generate over-exposure areas image with higher subjective quality, and the subjective quality of the sun and flame in the figure is improved.}
  \label{ablpdcg}
  \setlength{\belowcaptionskip}{-0.cm}
  \vspace{-15pt}
\end{figure}


\subsection{Experiment Results}

\textbf{Qualitative and Quantitative Results}. 
We first compare our method in Table.\ref{Quantitativetable} to compare with other methods in PSNR, SR-SIM, SSIM, $\Delta E_{ITP}$,HDR-VDP3.
PSNR (Peak Signal to Noise Ratio) can measure the pixel value difference between images.
SSIM (Structural Similarity Index Measure) evaluates the structural similarity of two images.
SR-SIM is able to evaluate the image similarity of HDR standard images.
$\Delta E_{ITP}$ can evaluate the colour difference of HDRTV.
HDR-VDP3 can assess image visual difference. Compared to HDR-VDP2, HDR-VDP3 supports the BT.2020 gamut.
The test results were calculated on 117 images with a resolution of 2160x3840. We calculate HDR-VDP3 scores on linear HDR images.
As Table.\ref{Quantitativetable}, our method produces significantly better objective results, which indicates the ability of our network to accurately reconstruct HDR frames.
To demonstrate the visual effect of the proposed method, we directly save the 16bit bit-depth image in PNG format, which is able to preserve all image information despite the fact that such a saving method will gray out the image.
Another method is to convert 16bit to 8bit using tone mapping, but this conversion removes some overexposed areas, so the result of generating overexposed areas between different methods cannot be shown.
The direct visualization method preserves all the details and the visualization results are displayed in Fig.\ref{Fig1} and Fig.\ref{QualitativeImage1}.
We also demonstrate in Fig.\ref{QualitativeImage1} that our proposed method can construct adaptive feature mappings for localities, thus mitigating to some extent the quality degradation caused by overexposure.
It can be seen that the previous method is unable to construct adaptive mapping for local images, and the HDCFM proposed in this paper is able to generate HDR frames with higher quality.

In addition, as shown in the lower right corner of Fig.\ref{QualitativeImage2}, HDCFM cannot perfectly recover the details of HDR frames when the input SDR image is overexposed due to tone mapping.
Nevertheless, our results are still an improvement compared to previous state-of-the-art methods.

To further analyze the performance of the proposed method, we calculated the histogram of the generated HDR video frames. 
The pixel intensity distribution of the HDR video frames generated by the proposed method is closer to the ground truth, and the pixel intensity distribution is smoother.
As shown in Table.\ref{Quantitativetable}, the proposed model HDCFM outperforms the past method approach in all evaluation metrics.
And the number of parameters of the HDCFM model is much smaller than that of the past method.

\begin{table}[t]
    \vspace{-5pt}
    \caption{Ablation study to verify the validity of the four modules, M0,M1,M2,M3,M4 refer to Global Feature Modulation, Local Linear Feature Transformation , Context Convolution, Local Feature Modulation, and Hierarchical Local Feature Modulation. The addition of each part can bring about the improvement of PSNR and other indicators, which proves that each module is indeed effective. }
    \centering
    \begin{spacing}{1.2}
    \begin{tabular}{m{0.4cm}<{\centering}m{0.4cm}<{\centering}m{0.4cm}<{\centering}m{0.4cm}<{\centering}m{0.4cm}<{\centering}m{0.7cm}<{\centering}m{0.72cm}<{\centering}m{0.72cm}<{\centering}m{1.0cm}<{\centering}} 
    \toprule
    \scriptsize M0    &  \scriptsize M1    & \scriptsize M2     & \scriptsize M3      & \scriptsize M4     & \scriptsize PSNR$\uparrow$ & \scriptsize SSIM$\uparrow$ & \scriptsize $\Delta E_{ITP}$$\downarrow$ & \scriptsize SR-SIM$\uparrow$ \\ 
    \midrule
    \Checkmark & \XSolidBrush    & \XSolidBrush     & \XSolidBrush    & \XSolidBrush     & 36.88 & 0.9655  & 9.78 &  0.9967 \\
    \Checkmark & \Checkmark & \XSolidBrush     & \XSolidBrush    & \XSolidBrush     & 37.74 & 0.9705 & 8.85 & 0.9972 \\  
    \Checkmark & \Checkmark & \Checkmark  & \XSolidBrush    & \XSolidBrush     & 38.20 & 0.9725 & 8.06 & 0.9972 \\ 
    \XSolidBrush & \Checkmark & \Checkmark  & \Checkmark & \XSolidBrush     & 38.26 & 0.9729 & 7.90 & 0.9973 \\   
    \Checkmark & \Checkmark & \Checkmark  & \Checkmark & \Checkmark     & 38.42 & 0.9732 & 7.83 & 0.9974 \\
    \bottomrule
    \end{tabular}
\end{spacing}
            \label{abl}
  \vspace{-15pt}
\end{table}

\textbf{Ablation Study}.
We performed ablation experiments on the whole model to demonstrate the effectiveness of each module of the proposed method.
Table \ref{abl} shows the performance of the SDRTV-to-HDRTV transformation after the addition of different modules.
M0, M1, M2, M3, M4 refer to Global Feature  Modulation , Local Feature Transformation , Context Convolution, Local Feature Modulation, and Hierarchical Local Feature Modulation.
With the addition of Local Feature Transformation, the objective quality is improved due to the new feature transformation method, which can model more complex color feature transformations, PSNR and SSIM are improved by 0.86 and 0.005, respectively, and $\Delta E_{ITP}$ is reduced by 0.93.
With the addition of Context Convolution, the model captures the remote context information to extract features, and the corresponding PSNR and SSIM are improved by 1.32 and 0.007, respectively, and $\Delta E_{ITP}$ is reduced by 1.72.
By using Local Feature modulation instead of Global Feature modulation, the model can generate local feature modulation vectors, allowing different feature modulations to be applied to different regions of the same image. This approach improves the PSNR and SSIM metrics by 1.38 and 0.0074, respectively, and reduces $\Delta E_{ITP}$ by 1.88.
After adding the combined local and global Hierarchical Local Feature Transform, the model can perform both global and local feature modulation, and the corresponding PSNR, SSIM is improved by 1.54 and 0.0077, respectively, and $\Delta E_{ITP}$ is reduced by 1.95.

The proposed PDCG module is capable of generating more realistic HDR reconstruction frames that are capable of generating subjective and comfortable images of over-exposure areas.
 In Fig.\ref{ablpdcg} we show the comparison images of the results generated by the proposed method. 
In areas where the content is saturated and over-exposed (the sun part and the flame part), PDCG can address the existing artifacts and overexposure problems. 
Therefore, HDCFM uses the overexposed content for feature mapping, and the resulting HDR frame still has overexposure. 
PDCG can dynamically restore HDR frames in overexposed areas, so as to obtain HDR frames with higher subjective quality.

 \section{Conclusion}

 For standard dynamic range (SDR) to high dynamic range(HDR) video.
 Previous methods performed global conversion of HDR frames without taking into account local information and the quality of HDR frames in overexposed areas during conversion. 
 In this paper, we proposed a two-stage SDRTV to HDRTV scheme to address these two problems.
 In the first stage, a feature mapping model is proposed. 
 Proposed method can perform non-consistent mapping for image local information, and the proposed dynamic feature transformation module is able to simulate more complex feature mapping. The converted HDR frames have a higher objective quality. 
 In the second stage, a patch discriminator and a context-based dynamic image generation model are constructed for overexposed areas. The patch discriminator can solve the problem that the model is difficult to train due to the low percentage of highlight areas. This model can improve the subjective quality of the reconstructed frames.
 Comprehensive experiments show that the proposed method achieves the best performance in both objective and subjective quality.

 \clearpage

\bibliographystyle{ACM-Reference-Format}
\bibliography{sample-base}

\end{document}